\documentclass{ws-ijmpd}

\begin{document}

\title{An assessment of the newest magnetar-SNR associations}

\author{J. E. Horvath $^{\star}$}
\author{M. P. Allen $^{\diamond}$}

\address{$\star$ Instituto de Astronomia, Geof\'\i sica e Ci\^encias
Atmosf\'ericas - Universidade de S\~ao Paulo\\
Rua do Mat\~ao, 1226, 05508-900, Cidade Universit\'aria, S\~ao
Paulo SP, Brazil}
\address{$\diamond$CEFET/SP, R. Pedro Vicente 625, Canind\'e, S\~ao Paulo SP, Brazil}

\maketitle

\begin{abstract}
Anomalous X-ray Pulsars and Soft-Gamma Repeaters groups are
magnetar candidates featuring low characteristic ages ($\tau =
{P\over{2 {\dot P}}}$). At least some of them they should still be
associated with the remnants of the explosive events in which they
were born, giving clues to the type of events leading to their
birth and the physics behind the apparent high value of the
magnetar magnetic fields. To explain the high values of $B$, a
self-consistent picture of field growth also suggests that energy
injection into the SNR is large and unavoidable, in contrast with
the evolution of {\it conventional} SNR. This modified dynamics,
in turn, has important implications for the proposed associations.
We show that this scenario yields low ages for the new candidates
CXOU J171405.7-381031/CTB 37B and XMMU
J173203.3-344518/G353.6-0.7, and predicted values agree with
recently found ${\dot P}$, giving support to the overall picture.
\end{abstract}

\section{Energy injection in magnetar-driven SNRs}

Several studies of the expansion of remnants in different ISM were
performed over the years, just assuming that the explosion is
successful an an energy of $\sim \, 10^{51} \, erg$ \cite{Mario}
is released in a point-like region and unrelated to the compact
remnant. Such vision was first challenged by Ostriker and Gunn
\cite{OG} when they postulated that the rotation energy of a
central pulsar may drive the supernova. Later it became clear that
this phenomenon is real but too slow to power explosions, although
the so-called magnetodynamical mechanisms \cite{Gena} still retain
many of these features.

The identification of superstrong magnetic fields in magnetars
posed yet a further related problem to the explosion scenarios. If
the field had to be amplified from an initial seed, then the
pre-supernova progenitors must have a suitable distribution of
both magnetic field and angular momentum. If the initial rotation
is not fast enough, the amplification of the field is quenched
\cite{Tho}. This leads to think that magnetars should be born
rotating very fast for their fields to grow, although their
braking is very efficient in the aftermath following their birth.
Thus, if a {\it magnetar} is formed inside the remnant immediately
after the explosion by dynamo amplification, the injection of
energy (much in the same way as Ostriker and Gunn 1971 envisioned)
is inevitable, and an initial energy loss $L_{0} = 3.85 \times
10^{47} {\biggl({\frac{B}{10^{14} \, G}}\biggr)}^{2}
{\biggl({\frac{1 \, ms}{P_{0}}}\biggr)}^{4} \, erg s^{-1}$ and
initial timescale for deceleration $\tau_{0} \, = \, 0.6
{\biggl({\frac{10^{14} \, G}{B}}\biggr)}^{2}
{\biggl({\frac{P_{0}}{1 \, ms}}\biggr)}^{2} \, d$ can be defined
for this process \cite{nos}.

Since the injected energy scales as $B^{-2}$ and the initial $ms$
periods are {\it required} for field amplification to operate, the
timescale for a substantial energy injection is in fact very
short, of the order of $\sim hours$ (a ``normal'' pulsar would
inject the same amount in $\sim 100-1000 yr$). A recent work
\cite{Woo} showed that a bright supernova and spectra would
follow, thus modified dynamical behavior at later times is
naturally expected.

The injected energy will make a very young remnant to expand {\it
more quickly} than the corresponding case without energy
injection, making it look older in the free-expansion phase
(lasting just $\sim 100 yr$) and later affecting the Sedov-Taylor
phase, when the internal energy of the gas inside the cavity $U$
picks a term

\begin{equation}\label{energy}
U \, = \, E - {\frac{9}{32}}M{\dot R}^{2} -
\frac{L_{0}}{t^{-1}+{\tau_{0}}^{-1}}
\end{equation}
\\
where $R$ the radius of the SNR and $M$ the mass in motion, and
the last term represents the injected energy due to the magnetar
formation. We refer to \cite{nos} for a thorough discussion of
these features. Comparison with works dealing with standard SNR
dynamics \cite{TMcKee} \cite{LuzBerry}) show that the Sedov-Taylor is
also modified to last longer than the case without injection, and
ends after $\sim 2 \times 10^{4} E_{51}^{3/14}n^{-4/7} \, yr$
\cite{nos}, when the SNR enters a ``snowplow'' (radiative) phase.
A specific search for these effects was undertaken in \cite{VK},
with negative results. However, we must remark that after $\sim \,
1000 \, yr$ the speed of the ejecta is essentially the same for
models with or without energy injection (Fig. 2 of Ref. [4]),
making a kinematical/spectroscopical identification more
difficult.

\section{The new candidate associations}

The analysis of the formerly proposed associations has been
carried by several groups \cite{nos} \cite{Gaensler} \cite{Ankay} \cite{Mardsen},
with varying results. Using the modified dynamics, we concluded in
\cite{nos} that the associations AXP 1845-0258/G29,6+0,1; AXP
2259+586/CTB 109 perhaps AXP 1709-4009/G346,5-0,1 among the AXP
and 1806-20/G10,0-0,3 ; SGR 1801-23 W 28 among the SGR passed this
test, although they some could and have been dismissed by other
reasons.

The suggestion of the newest associations CXOU
J171405.7-381031/CTB 37B and XMMU J173203.3-344518/G353.6-0.7 by
Halpern and Gotthelf \cite{HG} should be analyzed accordingly.
Although the period and $\dot{P}$ of XMMU J173203.3-344518 could
not be confirmed, Halpern and Gotthelf (2010b) have now reported a
strong field value for the former, qualifying it as a magnetar. A
full discussion of these SNRs has been addressed in several works
\cite{Aharonian1} \cite{Aharonian2} \cite{HG}. Fig 1 displays the Radius-Age
expected from SNRs with energy injection by a magnetar for two
extreme cases: a low-mass, low ISM density (labelled ``low-low'')
case of $M = 8 M_{\odot}$ and $n = 0.01 cm^{-3}$ explosion; and a
high-mass $M= 30 M_{\odot}$ and $n = 10 cm^{-3}$ one
(``high-high''), hopefully bracketing the range of
progenitor-ambient possibilities. The implied intervals of ages
from just this plot are $800-8000 yr$ for CXOU J171405.7-381031 in
CTB 37B and $1000-12000 yr$ for XMMU J173203.3-344518 in
G353.6-0.7. This may seem little restrictive, but firmly excludes
the estimate of \cite{Tian} among others, just as a result of the
modified dynamics discussed before. Since there is some
information about the density in the work of \cite{Aharonian2}, we
believe it is fair to consider a curve with $n = 0.5 cm^{-3}$, or
50 times bigger than the ``low-low'' curve constituting the upper
envelope. For the sake of comparison, the case of $M= 10
M_{\odot}$ and $n = 1 cm^{-3}$ is also plotted, if this were the
case the ages wold be further tightened to $2000 yr$ and $3000 yr$
respectively.

\begin{figure}[h!]
 \centering
  \includegraphics[scale= 0.8]{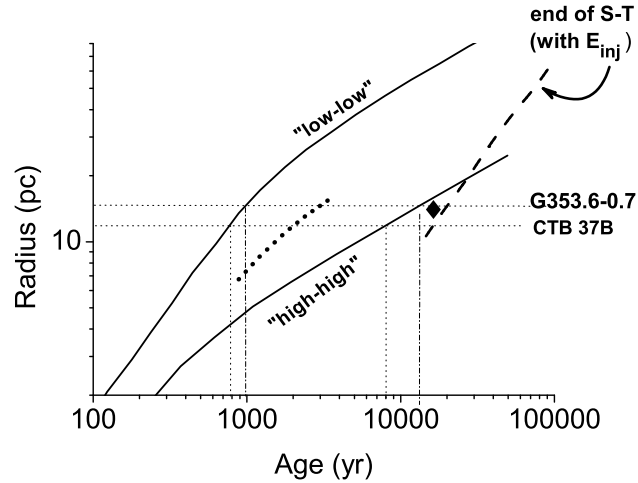}
  \caption{The ages of magnetar-driven SNR. The limiting
curves ``low-low'' and ``high-high'' are explained in the text.
The dotted line corresponds to the expansion of a $M= 10
M_{\odot}$ mass in a $n=1 cm^{-3}$ ambient. The end of the
Sedov-Taylor phase {\it locus} is diagonal dashed line on the
right of the figure. All the explosion energies were fixed to
$10^{51} erg$. Note that only very heavy progenitors
(``high-high'' or heavier) would produce the largest ages, making
$CTB 37B$ older than $10^{4} \, yr$.}
\label{FIGURA}
\end{figure}

\section{Discussion and Conclusions}

The magnetar progenitors remain largely unknown, and an initial
expectation of a high-mass ($> 30-40 M_{\odot}$) \cite{Brian}
derived from an interpretation of the HI shell GSH 288.3-0.5-28 as
a wind-bubble associated to the progenitor of 1E 1048.1-5937; and
later supported by the cluster analysis by \cite{Fig, Mun, Bib},
has been recently challenged by the identification of SGR 1900+14
by Davies et al. (2009) with the Cl 1900+14 cluster, implying a
progenitor of $< 17 M_{\odot}$. Those results suggest that the
difference between the events producing pulsars and magnetars
should not be just related to the mass of the progenitor. We
plotted in Fig. 1 the position of the shell GSH 288.3-0.5-28
interpreted as a ``standard'' (in the sense of the explosion
energy) SNR originated from a massive star ($\geq \, 30
M_{\odot}$) expanding in a dense environment (diamond symbol).
There is no difficulty with the energetics, and the inferred age is $\sim
15000 yr$, provided the modified dynamics is employed.

We have inferred ages for G353.6-0.7 and CTB 37B which are much
smaller than the values derived within conventional models.
Refining the broad interval to include a ISM density closer to
usual values (as directly measured for CTB 37B by
\cite{Aharonian2}), the figures for the ages are still low and
would not change much unless the mass of the progenitor was $\geq
4 \times 10 M_{\odot}$ (see Fig. 1). The age of this remnant is
closer to the ``old'' estimation of $\sim 1500 yr$ \cite{CS}, but
for quite different dynamical reasons. This age allowed a {\it
prediction} of the $\dot P$ value, provided the characteristic age
is close to the actual value, of $\dot P \, \sim \, 4 \times
10^{-11} s \, s^{-1}$, which in turn predicts a magnetic field
strength of $B = 4 \times 10^{14} G$, almost exactly the values
obtained by Halpern and Gotthelf \cite{HGott} for CXOU
J171405.7-381031 right after the Cesme event (!). It is also
interesting to note that younger objects easy the requirements for
energizing the TeV scale, as observed by the HESS Collaboration
\cite{Aharonian2}: electrons still ``live'' without being cooled
(\cite{HG}) or even the SNRs themselves contribute, because they
are actually younger than they seem.

To summarize, we believe that we are not dealing with ordinary
remnants, but rather with a very special variety of them, the
magnetar-driven ones, a natural result of dynamo field growth to
the $10^{14}-10^{15} G$ scale. It is in this framework that
associations need to be analyzed. An alternative picture in which
field values result from flux conservation would not render
substantial dipole energy injection.

\section*{Acknowledgments}
We acknowledge the financial support received from the Funda\c
c\~ao de Amparo \`a Pesquisa do Estado de S\~ao Paulo and CNPq
Agency (Brazil). JEH wishes to thank the hospitality of the
Organizers of ASTRONS2010.

\end{document}